\documentclass[12pt, prd, showpacs]{revtex4-1}
\usepackage{amssymb}
\usepackage{amsmath}
\usepackage{graphicx}

\begin{document}

\title{Redshift/blueshift inside the Schwarzschild black hole}
\author{O. B. Zaslavskii}
\affiliation{Department of Physics and Technology, Kharkov V.N. Karazin National
University, 4 Svoboda Square, Kharkov 61022, Ukraine}
\affiliation{Kazan Federal University, Kremlevskaya 18, Kazan 420008 Russia}
\email{zaslav@ukr.net}

\begin{abstract}
We consider an observer who moves under the horizon of the Schwarzschild
black hole and absorbs a photon. There are two different situations when (i)
a photon comes from infinity, (ii) it is emitted by another observer under
the horizon (say, by a surface of a collapsing star). We analyze the
frequency change for absorption near the event horizon, in an intermediate
region and near the singularity and compare the results for both scenarios.
Near the singularity, in both scenarios infinite redshift (in the pure
radial case) or infinite blueshift are possible, depending on angular
momenta of a photon and an observer. The main difference between both
scenarios manifests itself, if photons from the star surface are emitted
near the horizon and received in the intermediate region or near the
singularity by the observer with a nonzero angular momentum.
\end{abstract}

\maketitle

\section{Introduction}

Sometimes, strong prejudice against careful studies of properties of the
inner region of black holes reveals itself because of impossibility to
obtain information from there in "our" world (on the classical level).
Meanwhile, black holes is one of the most remarkable and sound predictions
of general relativity. They are also found in many other gravitation
theories. Thus the presence of a region under the event horizon is quite
solid physical result of theory. And, as a part of a physical world, it
deserves investigation in spite of some unusual properties of this region
(or, by contrary, just due to this fact). The situation in this area of
research was recently envisaged in a brief review \cite{ham2} where it was
stressed that because of neglect of the corresponding subject, even some
primary notions relevant for this region are described in literature not
quite accurately.

As far as the properties of the region inside a black hole is concerned, one
of the first questions is the view of surrounding world. What does a falling
observer see during his fall? In their popular book, Gurevich and Gliner
wrote that when the world line of an observer approaches the singularity, he
sees a surrounding world to fade (\cite{gg}, p. 59). However, they arrive at
this conclusion considering the value of the coordinate speed of light - the
quantity that does not have direct physical meaning. Quite recently, a
similar conclusion was made in \cite{ham} where propagation of light from
the "illusory horizon" was considered. This term was coined in \cite{ham} to
replace a more usual one "past horizon" in the complete analytically
extended Schwarzschild metric. Meanwhile, if, instead of the complete
space-time, realistic collapse of a star is considered, its surface follows
a time-like trajectory everywhere (including the inner region) and does not
approach the would-be past horizon, in contrast to what is supposed in \cite%
{ham} (especially, see page 13 and Fig. 6 there). Therefore, in our opinion,
the conclusions derived in \cite{ham} about exponential redshift \textit{%
inside} the event horizon, hang in the air. And, only radial motion of
particles and photons in the inner region was considered in \cite{ham}.

In the present work, we consider red/blue shift of light under the event
horizon in a general case and argue that account for nonzero angular momenta
of particles can change a whole picture drastically. In doing so, we
concentrate on the properties of a frequency near the singularity. The
results are valid both for an eternal black hole or for the realistic
collapse of a star outside its surface in the vacuum region. The formulas
for the frequency are very simple but, to the best of our knowledge, the
corresponding results that follow from them near the singularity were absent
from the literature. This work can be considered as a step towards more
general goal - constructing the whole picture seen by an observer inside a
black hole. This will have to include transformation of angles under which
light comes into view that deserves further separate study. In what follows,
we use the geometric system of units in which fundamental constants $G=c=1$.

\section{General equations}

\subsection{Metric}

Let us consider the black hole metric%
\begin{equation}
ds^{2}=-dt^{2}f+\frac{dr^{2}}{f}+r^{2}d\omega ^{2}\,\text{,}  \label{sph}
\end{equation}%
where $d\omega ^{2}=d\theta ^{2}+\sin ^{2}\theta d\phi ^{2}$, $f=f(r).$ The
roots of equation $f=0$ correspond to the horizon. We will discuss the case
when there is only one horizon $r_{+}\,\ $and mainly focus on the
Schwarzschild metric. Then, $f=1-\frac{r_{+}}{r}$, $r_{+}$ being the horizon
radius.

We are interested in the region inside the event horizon. As is known, the
metric can be described there by its original form (\ref{sph}) but with an
important reservation that spacelike and timelike coordinates mutually
interchange their roles - see \cite{nov1} or \cite{fn} (page 25).
Correspondingly, we redefine $t=y$, $r=-T$, $f=-g$, then the metric (\ref%
{sph}) takes the form%
\begin{equation}
ds^{2}=-\frac{dT^{2}}{g}+dy^{2}g+T^{2}d\omega ^{2}\text{.}  \label{met}
\end{equation}%
Here, all metric coefficients depend on $T$ only. For the Schwarzschild
metric,%
\begin{equation}
g=\frac{r_{+}}{(-T)}-1\text{, }-r_{+}\leq T\leq 0\text{.}  \label{g}
\end{equation}%
The hypersurface $T=const$ represents a hypercylinder extended in the $y$
direction. It is instructive to write down equations of geodesic motion for
massive particles and photons separately.

\subsection{Motion of massive particle}

As the metric does not depend on $y$ and $\phi $, the radial momentum $%
P\,=mu_{y}\ $and the angular one $\mathcal{L}=mu_{\phi }$ are conserved,
where $u^{\mu }=\frac{dx^{\mu }}{d\tau }$ is the four-velocity, $\tau $
being the proper time. Equations of motion within the plane $\theta =\frac{%
\pi }{2}$ for a geodesic particle read%
\begin{equation}
g\dot{y}=-p\text{,}  \label{my}
\end{equation}%
\begin{equation}
T^{2}\dot{\phi}=\mathcal{L}\text{, }  \label{mfy}
\end{equation}%
where $p=\frac{P}{m}$, $\mathcal{L=}\frac{L}{m}$, dot denotes derivative
with respect to the proper time $\tau $. If a particle having the energy $E>0
$, moves outside, the equation of motion gives $mf\dot{t}=E$ for it. When it
enters the inner region, $f=-g<0$, so for $\dot{y}$ we obtain just (\ref{my}%
) with $p=\frac{E}{m}$. However, if \ a particle inside did not come from
the outer region, $p$ can have any sign, $p=\pm \left\vert p\right\vert .$
Under the horizon, $p$ has the meaning of radial momentum and is still
conserved. The case $p=0$ is also possible \cite{lobo}. From the
normalization \ condition $u_{\mu }u^{\mu }=-1$ we have%
\begin{equation}
\frac{p^{2}}{g}+\frac{\mathcal{L}^{2}}{T^{2}}-\frac{\dot{T}^{2}}{g}=-1\,.
\end{equation}%
Taking into account the forward-in-time condition $\dot{T}>0$ we obtain

\begin{equation}
\dot{T}=Z\text{, }  \label{mT}
\end{equation}%
\begin{equation}
Z=\sqrt{p^{2}+g(\frac{\mathcal{L}^{2}}{T^{2}}+1)}\text{.}  \label{zt}
\end{equation}%
Thus in coordinates $(T,y,\phi )$ we have for the four-velocity $u^{\mu }=%
\frac{dx^{\mu }}{d\tau }$ 
\begin{equation}
u^{\mu }=(Z,-\frac{p}{g},\frac{\mathcal{L}}{T^{2}})\text{,}  \label{muu}
\end{equation}%
\begin{equation}
u_{\mu }=(-\frac{Z}{g},-p,\mathcal{L}).  \label{mul}
\end{equation}%
We omit $u^{\theta }=0$.

\subsection{Three-dimensional velocity}

It is instructive to define a velocity as a spatial three-dimensional vector 
$V^{(i)}$ ($i=1,2$). Actually, it is two-dimensional for motion within the
equatorial plane. To this end, we can define the tetrad (actually, there are
only 3 independent vectors since we consider the equatorial plane only but
we still use the standard term "tetrads" for convenience):%
\begin{equation}
h_{(0)\mu }=-\frac{1}{\sqrt{g}}(1,0,0)\text{,}  \label{tet0}
\end{equation}%
\begin{equation}
h_{(1)\mu }=\sqrt{g}(0,1,0)\text{,}
\end{equation}%
\begin{equation}
h_{(2)\mu }=\left\vert T\right\vert (0,0,1)\text{.}  \label{tet2}
\end{equation}%
Such a choice of tetrads corresponds to the observer with the four-velocity $%
U^{\mu }=h_{(0)}^{\mu }$ who remains at rest ($y=const$). It is interesting
that inside the horizon this is a geodesic observer \cite{lobo} that was
impossible outside, where an observer at rest is accelerated.

Then, the standard definition gives us the tetrad components 
\begin{equation}
V^{(i)}=-\frac{h_{(i)\mu }u^{\mu }}{h_{(0)\mu }u^{\mu }}\text{.}
\end{equation}

It is easy to obtain from (\ref{zt}), (\ref{muu}) and (\ref{tet0}) - (\ref%
{tet2}) that 
\begin{equation}
V^{(1)}=-\frac{p}{Z}\text{,}
\end{equation}%
\begin{equation}
V^{(2)}=\frac{\mathcal{L}\sqrt{g}}{\left\vert T\right\vert Z}\text{,}
\end{equation}%
\begin{equation}
\left\vert \vec{p}\right\vert =\frac{V\sqrt{g}}{\sqrt{1-V^{2}}}\text{.}
\label{pp}
\end{equation}%
Here $V\ =\sqrt{V^{(1)2}+V^{(2)2}}$, $\left\vert \vec{p}\right\vert =\sqrt{%
p^{2}+\frac{\mathcal{L}^{2}}{T^{2}}g\text{ }}$is the absolute value of the
momentum that takes into account both spatial components. Then, (\ref{pp})
has the meaning of the standard relation between the momentum and velocity,
redshifted due to the factor $\sqrt{g}$. One can also define the local
energy as $E_{loc}=-mu_{(0)}$. Then, it follows from (\ref{mT}), (\ref{zt})
and (\ref{tet0}) - (\ref{tet2}) that $E_{loc}=\sqrt{m^{2}+\frac{\left\vert 
\vec{P}\right\vert ^{2}}{g}}$. By substitution of (\ref{pp}), we obtain%
\begin{equation}
E_{loc}=\frac{m}{\sqrt{1-V^{2}}}\text{.}  \label{Et}
\end{equation}

As under the horizon the metric depends on time, the energy is not conserved.

Eqs. (\ref{pp}), (\ref{Et}) can be thought of as counterparts of the formula 
\begin{equation}
E=E_{loc}\sqrt{f}  \label{E}
\end{equation}%
for the energy $E$ outside the horizon \cite{k} (see eq. 15 there), now they
are valid inside.

It is seen from (\ref{pp}) that in the horizon limit, when $g\rightarrow 0$,
the velocity $V\rightarrow 1$, so it approaches the speed of light. This
applies to each branch of the horizon.

\subsection{Motion of photon}

In a similar way, the components $k_{y}\equiv -q$, $k_{\phi }\equiv l$ of
the wave vector $k^{\mu }$ are conserved inside the horizon. If a photon
came from the outer region, $\left\vert q\right\vert =\omega _{0}$ has the
meaning of the frequency emitted by a static observer at infinity. But in
general, it can also be emitted already under the horizon.

The normalization condition $k_{\mu }k^{\mu }=0$ gives us%
\begin{equation}
-\frac{(k^{0})^{2}}{g}+\frac{l^{2}}{T^{2}}+\frac{q^{2}}{g}=0\text{,}
\end{equation}%
whence%
\begin{equation}
k^{0}=z\text{,}
\end{equation}%
where 
\begin{equation}
z=\sqrt{q^{2}+\frac{g}{T^{2}}l^{2}}\text{.}  \label{z}
\end{equation}%
Then, the wave vector is equal to%
\begin{equation}
k^{\mu }=(z,-\frac{q}{g},\frac{l}{T^{2}})\text{,}  \label{k}
\end{equation}%
\begin{equation}
k_{\mu }=(-\frac{z}{g},-q,l)\text{.}
\end{equation}

We remind a reader that we consider motion in the equatorial plane only, so
all relevant momenta have no $\theta $ component.

\section{Frequency}

The frequency measured by an observer with the four-velocity $u^{\mu }$ is
equal to%
\begin{equation}
\omega =-u_{\mu }k^{\mu }\text{.}
\end{equation}%
Taking into account eqs. (\ref{mul}) and (\ref{k}), we have%
\begin{equation}
\omega =\frac{zZ-pq}{g}-\frac{\mathcal{L}l}{T^{2}}.  \label{om}
\end{equation}

Our main concern is the behavior of the frequency near the horizon and
singularity. The relevant quantities enter the general expression (\ref{om})
in such a way that, as a rule, smooth limiting transitions to the particular
cases are impossible. For instance, in the combination $\frac{\mathcal{L}^{2}%
}{T^{2}}g$ in (\ref{zt}) the result depends strongly on which limit is taken
first - $\mathcal{L}\rightarrow 0$ or $T\rightarrow 0,$ etc. Therefore, we
will consider some particular physically interesting situations separately,
case by case.

\section{Exact formulas for particular cases}

To facilitate reading, we give at first explicitly general exact formulas in
particular cases, even in spite of their simplicity. Afterwards, we will
analyze them near the horizon and singularity.

\subsection{Radial motion of a photon: $l=0$, $\mathcal{L}\neq 0$}

It follows from (\ref{z}) and (\ref{om}) that

\begin{equation}
\omega =\frac{\left\vert q\right\vert (Z-\alpha \left\vert p\right\vert )}{g}%
\text{,}  \label{rad}
\end{equation}%
\begin{equation}
\alpha =sign(pq),  \label{al}
\end{equation}%
$Z$ is given by (\ref{zt}).

\subsection{Radial motion of an observer: $\mathcal{L}=0$, $l\neq 0$.}

Then, we have from (\ref{zt}), (\ref{om})

\begin{equation}
\omega =\frac{\sqrt{(p^{2}+g)(q^{2}+\frac{l^{2}}{T^{2}}g)}-pq}{g}\text{.}
\label{L0l}
\end{equation}

\subsection{Radial motion of an observer and a photon: $\mathcal{L}=0$, $l=0$%
.}

Let both $\mathcal{L}=0$, $l=0$. Then, it follows from (\ref{rad}) that%
\begin{equation}
\omega =\frac{\left\vert q\right\vert (\sqrt{p^{2}+g}-\alpha \left\vert
p\right\vert )}{g}=\frac{\left\vert q\right\vert }{\sqrt{p^{2}+g}+\alpha
\left\vert p\right\vert }.  \label{Zp}
\end{equation}

In this case, it is seen from (\ref{rad}), (\ref{Zp}) that for both signs of 
$\alpha $,%
\begin{equation}
\frac{d\omega }{dg}<0\text{.}
\end{equation}

In the Schwarzschild metric $\frac{dg}{dr}<0$ everywhere under the horizon,
so $\frac{d\omega }{dr}>0$. When a particle moves under the horizon towards
the singularity $r=0$, $\omega $ diminishes, so the redshift is increasing
in the process of motion.

\subsection{Angular motion of a photon: $q=0$}

The limit $q\rightarrow 0$ corresponds to a photon that does not move in the
"radial" direction along the leg of a hypercylinder and only circumscribes
the half of a full circle in the angular direction (see eq. 69 of \cite{lobo}%
). If $q=0$ it is necessary that $l\neq 0$ to have non-vanishing $k^{\mu }$.

Then, for $\mathcal{L}\neq 0$, it follows from (\ref{om}) that%
\begin{equation}
\omega =\frac{\left\vert l\right\vert }{T}\left( \sqrt{\frac{\mathcal{L}^{2}%
}{T^{2}}+1+\frac{p^{2}}{g}}-\frac{\left\vert \mathcal{L}\right\vert sgn%
\mathcal{L}l}{\left\vert T\right\vert }\right) \text{.}  \label{Q0}
\end{equation}

For $\mathcal{L}=0$, we can obtain from (\ref{Q0})%
\begin{equation}
\omega =\left\vert l\right\vert \frac{\sqrt{p^{2}+g}}{\sqrt{g}\left\vert
T\right\vert }.  \label{orad}
\end{equation}

\subsection{Observer at rest in y direction: $p=0$}

Under the horizon, a geodesic observer can have $p=0$ and even remain at
rest in the corresponding frame (\ref{met}) if $\mathcal{L}=0$ as well (see
Sec. 2.2. of \cite{lobo} for details of such a trajectory). This property
has no analog outside the horizon since the radial momentum depends on time
in the outer region but it is conserved in the inner one. Now, we have from (%
\ref{zt}), (\ref{z}), (\ref{om})%
\begin{equation}
\omega =\frac{\sqrt{\frac{\mathcal{L}^{2}}{T^{2}}+1}\sqrt{q^{2}+\frac{l^{2}g%
}{T^{2}}}}{\sqrt{g}}-\frac{\mathcal{L}l}{T^{2}}\text{.}  \label{P}
\end{equation}

If $\mathcal{L}=0$, $l\neq 0$,%
\begin{equation}
\omega =\frac{\sqrt{q^{2}+\frac{l^{2}g}{T^{2}}}}{\sqrt{g}}\text{.}
\label{PL}
\end{equation}

If $\mathcal{L}\neq 0$, $l=0$,%
\begin{equation}
\omega =\frac{\sqrt{\frac{\mathcal{L}^{2}}{T^{2}}+1}\left\vert q\right\vert 
}{\sqrt{g}}\text{.}  \label{Pl0}
\end{equation}

If $\mathcal{L}=0=l$,%
\begin{equation}
\omega =\frac{\left\vert q\right\vert }{\sqrt{g}}\text{.}  \label{Ll}
\end{equation}%
If $p=q=0$, we see from (\ref{P}) that%
\begin{equation}
\omega =\frac{\sqrt{\frac{\mathcal{L}^{2}}{T^{2}}+1}\left\vert l\right\vert 
}{\left\vert T\right\vert }-\frac{\mathcal{L}l}{T^{2}}\text{.}  \label{PQ}
\end{equation}

If $p=q=\mathcal{L}=0$, $l\neq 0$, it follows from (\ref{L0l}) that 
\begin{equation}
\omega =\left\vert \frac{l}{T}\right\vert \text{.}  \label{lt}
\end{equation}

\bigskip

One can check that all these formulas are mutually consistent with each
other. For example, if we put $p=0$ and $\mathcal{L}=0$ in (\ref{Q0}) or $%
\mathcal{L}=0$ in (\ref{PQ}) we obtain the same result (\ref{lt}), etc.

Now, on the basis of the obtained formulas, we analyzed behavior of the
frequency near the horizon.

\section{Photon absorbed near the horizon}

Now, we assume that $g\rightarrow 0$.

\subsection{Generic case}

If $\alpha =+1,$ we have from (\ref{om})

\begin{equation}
\omega (r_{+})=\frac{q}{2p}+\frac{(pl-q\mathcal{L})^{2}}{2pqr_{+}^{2}}.
\label{gen}
\end{equation}%
Eq. (\ref{gen}) is valid for generic $\mathcal{L}$, $l.$ In the particular
case $\mathcal{L}=0$ it agrees with eq. (10) of \cite{along}. For the pure
radial case, $\mathcal{L}=l=0$,

\begin{equation}
\omega (r_{+})=\frac{q}{2p}\text{.}  \label{12}
\end{equation}%
If a particle crosses the horizon moving from infinity where it was at rest, 
$p=1.$ Taking also into account that the integral of motion $q$ has the
meaning of frequency at infinity, $q=\omega _{0}$, we have 
\begin{equation}
\frac{\omega (r_{+})}{\omega _{0}}=\frac{1}{2}.  \label{05}
\end{equation}

If $\alpha =-1$, eq. (\ref{om}) gives us 
\begin{equation}
\omega \approx \frac{2\left\vert pq\right\vert }{g}  \label{0pq}
\end{equation}%
independently of $\mathcal{L}$ and $l$, so $\omega \rightarrow \infty $ when
a photon is absorbed near the horizon. This is a kind of head-on collision.
However, such a collision with finite nonzero $p$ and $q$ (say, $p>0$ and $%
q<0$) near the horizon implies that in the complete Schwarzschild space-time
a photon crossed the horizon from the "mirror" universe. The similar effect
in the outer region occurs near the white hole horizon \cite{white}.
Meanwhile, in a more physical situation, if it is emitted from the surface
of a collapsing star, the requirement of the finiteness of $\omega $ entails
that $q$ itself is small (for more details see \ below).

\subsection{Angular motion of a photon: $q=0$, $p\neq 0$}

It follows from (\ref{Q0}) that 
\begin{equation}
\omega \approx \left\vert l\right\vert \frac{\left\vert p\right\vert }{\sqrt{%
g}r_{+}}  \label{Q=0}
\end{equation}%
independently of $\left\vert \mathcal{L}\right\vert $. Thus $\omega
\rightarrow \infty $ in the horizon limit $g\rightarrow 0$. If a massive
particle has zero momentum in $y$ direction, it passes though the
bifurcation point \cite{inner}. The situation when an observer has $p>0$,
crosses the horizon and meets there a photon with $q=0$ is a counterpart of
the situation considered in \cite{zero}, where $p=0$, $q\neq 0$.

\subsection{Observer at rest in y direction: $p=0$, $q\neq 0$}

It follows from (\ref{P}) that for any $l$

\begin{equation}
\omega \approx \frac{\sqrt{\frac{\mathcal{L}^{2}}{r_{+}^{2}}+1}\left\vert
q\right\vert }{\sqrt{g}}
\end{equation}%
diverges in the horizon limit $g\rightarrow 0$, so there is an infinite
blueshift. Such a high energy collision \cite{zero} can be considered as
some analogue of the BSW effect \cite{ban}.

\subsection{$q=0$, $p=0$}

Then, we have from (\ref{PQ}) that%
\begin{equation}
\omega (r_{+})=\frac{\sqrt{\frac{\mathcal{L}^{2}}{r_{+}^{2}}+1}\left\vert
l\right\vert }{r_{+}}-\frac{\mathcal{L}l}{r_{+}^{2}}  \label{pq0}
\end{equation}%
is finite and nonzero.

We can summarize the results in the table.

\begin{tabular}{|l|l|}
\hline
& $\omega $ \\ \hline
$pq>0$ & finite \\ \hline
$pq<0$ & infinite \\ \hline
$q=0$, $p\neq 0$ & infinite \\ \hline
$p=0$, $q\neq 0$ & infinite \\ \hline
$p=q=0$ & finite \\ \hline
\end{tabular}

Table 1. Behavior of the frequency near the horizon.

\section{Behavior near the singularity}

Near the singularity, $g\rightarrow \infty $, $r\rightarrow 0$, $%
T\rightarrow 0$.

The result depends strongly on angular momenta of an observer and a photon.

If $\mathcal{L}l>0$, we obtain from (\ref{om}) that%
\begin{equation}
\omega \approx \frac{l}{2\mathcal{L}}<\infty \text{.}  \label{lpos}
\end{equation}

$\mathcal{L}l<0$%
\begin{equation}
\omega \approx \frac{2\left\vert \mathcal{L}l\right\vert }{T^{2}}\rightarrow
\infty  \label{lneg}
\end{equation}

$\mathcal{L}=0$, $l\neq 0$

We have from (\ref{L0l}) that%
\begin{equation}
\omega \approx \left\vert \frac{l}{T}\right\vert \rightarrow \infty
\label{0l}
\end{equation}%
Eqs. (\ref{lpos}) - (\ref{0l}) are insensitive to $p$ and $q$.

$\mathcal{L}\neq 0$, $l=0$

From (\ref{zt}), (\ref{rad}) we obtain that

\begin{equation}
\omega \approx \frac{\left\vert q\right\vert \left\vert \mathcal{L}%
\right\vert }{\left\vert T\right\vert \sqrt{g}}.  \label{L0}
\end{equation}%
In the Schwarzschild case, $g\approx \left\vert \frac{r_{+}}{T}\right\vert $%
, so%
\begin{equation}
\omega \approx \frac{\left\vert q\right\vert \left\vert \mathcal{L}%
\right\vert }{\sqrt{\left\vert T\right\vert }\sqrt{r_{+}}}  \label{s}
\end{equation}%
diverges when $T\rightarrow 0$, $g\rightarrow \infty $.

$\mathcal{L}=0$, $l=0$

It follows from (\ref{Zp}) that 
\begin{equation}
\omega \approx \frac{\left\vert q\right\vert }{\sqrt{g}}\rightarrow 0.
\label{00}
\end{equation}

The results are summarized in the table. Here, the values of $p$ and $q$ and
their relative sign are irrelevant.

\begin{tabular}{|l|l|}
\hline
& $\omega $ \\ \hline
$\mathcal{L}l>0$ & finite nonzero \\ \hline
$\mathcal{L}l<0$ & infinite blueshift \\ \hline
$\mathcal{L}=0$, $l\neq 0$ & infinite blueshift \\ \hline
$\mathcal{L}\neq 0$, $l=0$ & infinite blueshift \\ \hline
$\mathcal{L}=0=l$ & infinite redshift \\ \hline
\end{tabular}

Table 2. Behavior of the frequency near the singularity

It is worth noting that although in case $\mathcal{L}l>0$ the frequency is
finite and nonzero, it can take any value depending on parameters. In
particular, if a photon was emitted with the frequency $\omega _{1}$ and
absorbed with the frequency $\omega _{2}$, both limiting cases $\omega
_{2}\ll \omega _{1}$ and $\omega _{2}\gg $ $\omega _{1}$ are possible.

\section{What will a falling observer see? Typical case \label{astr} }

Up to now, we considered the process of absorption of a photon by an
observer with given $\mathcal{L},p,q,$ not specifying a precedent act of
emission in which a photon with these characteristics appeared. If a photon
enters the region under the horizon from the outside, $q>0$ and the quantity 
$q$ is equal to the frequency at infinity $\omega _{0}.$ Let us call it
scenario (i).

Meanwhile, there is another scenario (ii). It has a two step character. At
first, a photon is emitted by observer 1 already under the horizon. Further,
it is received by observer 2. In doing so, the angular momentum $l$ of a
photon does not change during the travel between two events. We assume that
observer 2 falls from infinity and has $p_{2}>0$. The role of observer 1 can
be played, say, by a star collapsing surface (then, $p_{1}>0$) or some
particle from dust cloud that is able to radiate. We also assume that a
photon is sent to meet observer 2, so absorption has a kind of head-on
collision: $p_{2}>0$, $q<0$. Correspondingly, in eq. (\ref{al}) 
\begin{equation}
\alpha =-1  \label{al-1}
\end{equation}%
for both observers 1 and 2.

The space-time diagram that depicts both scenario, is presented in Fig. 1.
It is nothing else than a standard diagram describing a gravitation
collapse. On this diagram, line H corresponds to the horizon.
\begin{figure}[htb]
\center
\includegraphics[width=8cm,height=8cm]{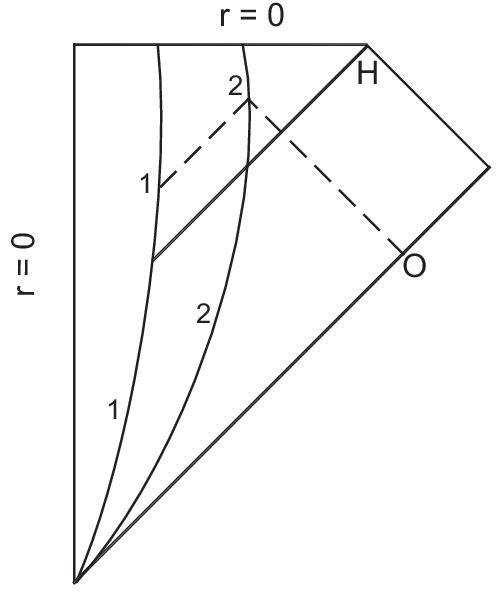}\hfill %
\caption{\label{Figs} Space-time diagram describing a star surface (line 1) and a
falling observer (line 2).}
\end{figure}

The most interesting region is the vicinity of the singularity. Meanwhile,
the results described above tell us that near the singularity the frequency
is insensitive to the sign of $q$, so the difference between scenarios (i)
and (ii) is blurred. And, from Table 2 we see that a quite diverse set of
situations becomes possible in each of two scenarios. 

While parameter $q$ becomes irrelevant near the singularity, the crucial
role is played by both angular momenta $\mathcal{L}$ and $l$. As a result,
typically either an unbounded blueshift or finite frequency shift can occur.
Only in the exceptional case of radial photon and radial observer, the final
outcome is an unbounded redshift in agreement wth \cite{gg}, \cite{ham}.

Another physically relevant region is the vicinity of the horizon. Let, for
simplicity, a signal is emitted by a source that is at rest at infinity.
Then, there is a universal relation (\ref{05}). Meanwhile, there is no such
a relation if a photon is emitted by a star surface. It was shown earlier
that, if an emitter crossed the horizon in a point with the value $V=V_{1}$
of the standard Kruskal coordinate and a receiver did it in the point $%
V=V_{2}$,%
\begin{equation}
\frac{\omega _{2}}{\omega _{1}}=\frac{V_{1}}{V_{2}}\text{.}  \label{omv}
\end{equation}%
This was shown for the Schwarzschild metric in \cite{kass} and in a more
general setting in \cite{along}. In doing so, a photon propagates exactly
along the horizon, $l=0=q$ (below, we enlarge these observations mentioned
in \cite{along}). Thus, in case (ii) the answer is not as universal as (\ref%
{05}).

\section{Signals emitted under the horizon: special cases of scenario ii}

In a previous section \ref{astr}, we discussed quite generic situations.
Meanwhile, there exist special cases that need an additional care for the
analysis in the framework of scenario (ii). In particular, this includes
case $p_{1}=0$. It does not correspond to a usual gravitational collapse of
a star but, for generality, we will consider this "exotic" example as well.
However, we omit from consideration case $p_{1}<0$. In the picture of an
external black hole it would correspond to a particle coming from the left
("mirror") universe but such a region is absent in the case of collapse of
matter. 

Another special case that requires additional attention implies the
following configuration. A photon is emitted in point 1 very close to the
horizon. If it propagates exactly along the horizon, it will be received in
point 2 also on the horizon with some finite redshift \cite{along}. If
points 1 or 2 somewhat change slightly, the frequency $\omega _{2}$ also
changes slightly. However, it may happen that points 1 and 2 are separated
largely, so that $V_{2}\gg V_{1}$. Then, according to (\ref{omv}), a
redshift can be made as strong as one likes. What is also important, a
photon emitted very close to the horizon, in such a situation can be
received in some intermediate point or even near the singularity since
observer 2 can hit line $r=0$ close to the upper right corner on Fig. 1. 

In doing so, new interesting options can appear. Let, for example, a surface
of a collapsing star radiate pure radial photons, $l=0$. Assuming that an
observer has $\mathcal{L}_{2}\neq 0$ and taking into account Table 2, we
come to the conclusion that in the course of his travel, the behavior of
frequency $\omega _{2}$ changes radically. There is a strong redshfit near
the horizon and in an intermediate region (see below) but, as the
singularity is approached, this changes to unbounded blueshift! 

Below we describe corresponding behavior of frequency in more detail and
derive restrictions on parameters of a photon emitted near the horizon that
generalize those in \cite{along}.

\subsection{Angular momentum of a photon emitted close to the horizon,
\thinspace p$_{1}>0$}

If an observer crosses the horizon and emits a photon in the direction
inside the horizon, $\alpha =+1$ and there is nothing special in this
process. Instead, if (\ref{al-1}) is fulfilled, there are severe
restrictions. Indeed, let in point of emission 1 the metric function $g_{1}$
be very small, formally $g_{1}\rightarrow 0$. We also assume that $p_{1}\neq
0$. For any physically reasonable process with a finite frequency $\omega
_{1}$ it is seen from (\ref{om}) that in this limit%
\begin{equation}
\sqrt{q^{2}+\frac{l^{2}}{r_{+}^{2}}g_{1}}+\left\vert q\right\vert \approx 
\frac{\omega _{1}}{p_{1}}g_{1}.  \label{gq}
\end{equation}

Bearing in mind that both terms in the left hand side are positive (or, at
least, nonnegative) we see that for any finite and nonzero $\omega _{1}$, $%
p_{1}$, scenario with $l\neq 0$ is impossible since it implies different
powers of $g_{1}$ in both sides of the equation. Eq. (\ref{gq}) is
self-consistent, if%
\begin{equation}
\left\vert q\right\vert \approx q_{1}g_{1}\text{, }l\approx l_{1}\sqrt{g_{1}}%
\text{,}  \label{q1}
\end{equation}%
where $q_{1}$ and $l_{1}$ are some constants. Then, it is seen from (\ref{gq}%
)\ that these parameters should satisfy the restriction%
\begin{equation}
\sqrt{q_{1}^{2}+\frac{l_{1}^{2}}{r_{+}^{2}}}+q_{1}=\frac{\omega _{1}}{p_{1}}%
\text{.}  \label{qpl}
\end{equation}

Thus in the limit under discussion the necessary condition reads $%
l\rightarrow 0$, $q\rightarrow 0$. If, for simplicity, we consider the case $%
l_{1}=0$, we see from (\ref{q1}), (\ref{qpl}) that 
\begin{equation}
\left\vert q\right\vert \approx \frac{\omega _{1}g_{1}}{2p_{1}}\text{.}
\label{p1}
\end{equation}

If a photon with $\alpha =-1$ is emitted just at the moment when an observer
crosses the horizon ($g_{1}=0$), $l=0$ exactly. Then, such a photon is
propagates along the horizon (see \cite{along} for details). Meanwhile,
another photon (with $\alpha =+1$) can have any $l$ and moves towards a
future singularity inside a black hole.

\subsection{Photon received in some intermediate point, p$_{1}>0$}

According to explanations above, for a photon with $\alpha =-1$, emitted
near the horizon, $l\approx 0$, so our photon moves almost radially.
Assuming $l_{1}=0$, we can consider it to be pure radial. Then, we can apply
eq. (\ref{0pq}), so for a finite $\omega _{1}\neq 0$ in the frame comoving
with observer 1 we have eq. (\ref{p1}). Here, the motion of the emitter can
be radial or not. The quantity $g_{1}$ is very small since by assumption
point 1 is near the horizon, so $q$ is small as well. Further, this photon
is received in point 2 where $g_{2}=O(1)$. Let us find its frequency $\omega
_{2}$.

Let $\mathcal{L}_{2}\neq 0$. From (\ref{rad}) and (\ref{p1}) we have%
\begin{equation}
\frac{\omega _{2}}{\omega _{1}}\approx \frac{g_{1}}{2p_{1}g_{2}}%
(Z_{2}+\left\vert p_{2})\right\vert \text{.}  \label{intl0}
\end{equation}

Now, according to (\ref{intl0}), $\omega _{2}\rightarrow 0$ since $%
g_{1}\rightarrow 0$. To obtain the case $\mathcal{L}_{2}=0$, we can take
safely the limit $\mathcal{L}_{2}\rightarrow 0$ in (\ref{intl0}), so%
\begin{equation}
\frac{\omega _{2}}{\omega _{1}}\approx \frac{g_{1}}{2p_{1}g_{2}}(\sqrt{%
p_{2}^{2}+g_{2}}+\left\vert p_{2})\right\vert \text{.}  \label{00int}
\end{equation}

\subsection{Angular momentum of a photon emitted close to the horizon,
\thinspace p$_{1}=0$}

The situation is different, if $p_{1}=0$. Then, it follows from (\ref{P})
near the horizon, where $g_{1}\rightarrow 0$, 
\begin{equation}
\sqrt{q^{2}+\frac{l^{2}g_{1}}{r_{+}^{2}}}\approx \frac{\left( \omega _{1}+%
\frac{\mathcal{L}_{1}l}{r_{+}^{2}}\right) }{\sqrt{\frac{\mathcal{L}_{1}^{2}}{%
r_{+}^{2}}+1}}\sqrt{g_{1}}\text{.}  \label{qp0}
\end{equation}

If $q=O(1)$ is separated from zero, the left hand side remains finite
nonzero, so for any finite $\omega _{1},$ eq. (\ref{qp0}) cannot hold. The
only way out is to assume that now%
\begin{equation}
\left\vert q\right\vert \approx q_{1}\sqrt{g_{1}}\text{.}  \label{qrg}
\end{equation}

Then, we obtain the constraint%
\begin{equation}
\sqrt{q_{1}^{2}+\frac{l^{2}}{r_{+}^{2}}}=\frac{\left( \omega _{1}+\frac{%
\mathcal{L}_{1}l}{r_{+}^{2}}\right) }{\sqrt{\frac{\mathcal{L}_{1}^{2}}{%
r_{+}^{2}}+1}}\text{.}  \label{ql}
\end{equation}%
In doing so, both cases $l=0$ or $l\neq 0$ separated from zero are allowed.
If $l=0$, the expression simplifies to%
\begin{equation}
q_{1}=\frac{\omega _{1}}{\sqrt{\frac{\mathcal{L}_{1}^{2}}{r_{+}^{2}}+1}}%
\text{.}  \label{q10}
\end{equation}

Thus there are two possible cases. If $p_{1}>0$, an observer, according to (%
\ref{q1}), can emit a photon near the horizon with a very small $l$ only. If
it is emitted just at the moment when an observer crosses the horizon, $l=0$
exactly. Then, such a photon propagates along the horizon (another photon
moves into the inner black hole region). This is the situation described
above and in \cite{along}.

If $p_{1}=0$, such an observer can pass through the horizon in the
bifurcation point only \cite{inner}. If, \ additionally, $l\neq 0$, a photon
cannot remain on the horizon, so both photons emitted move inside. If
emission occurs exactly on the horizon, $g_{1}=0$ there, so according to (%
\ref{qrg}), $q=0$ exactly.

\subsection{Photon received in some intermediate point, p$_{1}=0$}

If $l=0$, it follows from (\ref{rad}) that for $g_{1}\rightarrow 0$%
\begin{equation}
\frac{\omega _{2}}{\omega _{1}}\approx \frac{\sqrt{g_{1}}(Z_{2}+p_{2})}{%
\sqrt{\frac{\mathcal{L}_{1}^{2}}{r_{+}^{2}}+1}\sqrt{g_{2}}},  \label{p0}
\end{equation}%
where for simplicity we assumed that $l=0$ exactly.

Comparing (\ref{00int}) and (\ref{p0}) we see that the rate with which $%
\omega _{2}$ decreases when $g_{1}\rightarrow 0$, is higher (proportional to 
$g_{1}$) for $p_{1}>0$ than for $p_{1}=0$ (proportional to $\sqrt{g_{1}}$).

\subsection{Absorption near the singularity, $p_{1}>0$}

As is explained above, for physically reasonable emission near the horizon, $%
l=0$ or is negligible. Let now $\mathcal{L}_{2}\neq 0$. Then, eqs. (\ref{L0}%
), (\ref{s}), (\ref{p1}) apply,

\begin{equation}
\frac{\omega _{2}}{\omega _{1}}\approx \frac{g_{1}}{2p_{1}}\frac{\left\vert 
\mathcal{L}_{2}\right\vert }{\left\vert T_{2}\right\vert \sqrt{g_{2}}}.
\label{2}
\end{equation}

For the Schwarzschild case,%
\begin{equation}
\frac{\omega _{2}}{\omega _{1}}\approx \frac{g_{1}}{2p_{1}}\frac{\left\vert 
\mathcal{L}_{2}\right\vert }{\sqrt{\left\vert T_{2}\right\vert }\sqrt{r_{+}}}%
=\frac{\left\vert \mathcal{L}_{2}\right\vert \delta }{2p_{1}r_{+}},
\label{del}
\end{equation}%
$\delta =\frac{g_{1}\sqrt{r_{+}}}{\sqrt{\left\vert T_{2}\right\vert }}$.
Thus we have the play of two small quantities $g_{1}$ and $T_{2}$. If $%
\delta \ll 1$, $\frac{\omega _{2}}{\omega _{1}}\ll 1$ (strong redshift).
When $\delta \gg 1$, we have a strong blueshift. When $\delta =O(1)$, $%
\omega _{2}\neq 0$ is finite$.$ Eventually, as the singularity is
approached, $\delta \rightarrow \infty $, so blueshift prevails. Thus we see
the crucial change of the frequency from a big redshift near the horizon to
a strong blueshift near the singularity. The transition occurs for 
\begin{equation}
\left\vert T_{2}\right\vert =r_{2}\sim \left( \frac{g_{1}\left\vert \mathcal{%
L}_{2}\right\vert }{p_{1}}\right) ^{2}\frac{1}{r_{+}}.
\end{equation}

If $\mathcal{L}_{2}=0$, it follows from eq. (\ref{Ll}) that%
\begin{equation}
\frac{\omega _{2}}{\omega _{1}}\approx \frac{g_{1}}{2\sqrt{g_{2}}p_{1}}%
\approx \frac{g_{1}\sqrt{\left\vert T_{2}\right\vert }}{2\sqrt{r_{+}}p_{1}}%
\rightarrow 0  \label{ll00}
\end{equation}%
due to small $g_{1}$ and big $g_{2}$ (small $\left\vert T_{2}\right\vert $).

\subsection{Absorption near the singularity, $p_{1}=0$}

If $l=0$, $\mathcal{L}_{2}\neq 0$, eqs. (\ref{s}) and (\ref{q10}) gives us%
\begin{equation}
\frac{\omega _{2}}{\omega _{1}}\approx \frac{\sqrt{g_{1}}\left\vert \mathcal{%
L}_{2}\right\vert \sqrt{r_{+}}}{\sqrt{\left\vert T_{2}\right\vert }\sqrt{%
\mathcal{L}_{1}^{2}+r_{+}^{2}}}\text{.}
\end{equation}%
This shows that an infinite blueshift occurs in the limit $T_{2}\rightarrow 0
$. If $\,l=0$, $\mathcal{L}_{2}=0$,\ eq. (\ref{00}) is valid. Takinga lso
into account (\ref{qrg}), (\ref{q10}) we find%
\begin{equation}
\frac{\omega _{2}}{\omega _{1}}\approx \frac{\sqrt{g_{1}}}{\sqrt{g_{2}}\sqrt{%
\frac{\mathcal{L}_{1}^{2}}{r_{+}^{2}}+1}}\approx \frac{\sqrt{g_{1}}\sqrt{%
\left\vert T_{2}\right\vert }}{\sqrt{\mathcal{L}_{1}^{2}+r_{+}^{2}}}\text{.}
\end{equation}%
Eventually, the frequency $\omega _{2}\rightarrow 0$, when $g_{2}\rightarrow
\infty $. 

Now, let $l\neq 0$.  Then, if $\mathcal{L}_{2}=0$, eq. (\ref{0l}) applies,
so independently of $p_{2}$, there is an infinite blueshift here. If $%
\mathcal{L}_{2}\neq 0$, we have eqs. (\ref{lpos}) with finite $\omega _{2}$
or (\ref{lneg}) with an infinite blueshift depending on the sign of $%
\mathcal{L}l$.

It is convenient to collect the results in Table 4, where only dependence on
coordinate $T_{2}\rightarrow 0$ is shown.

\begin{tabular}{|l|l|l|}
\hline
& $p_{1}>0$ & $p_{1}=0$ \\ \hline
$l=0$, $\mathcal{L}_{2}\neq 0$ & $\frac{1}{\sqrt{\left\vert T_{2}\right\vert 
}}$ & $\frac{1}{\sqrt{\left\vert T_{2}\right\vert }}$ \\ \hline
$l=0$, $\mathcal{L}_{2}=0$ & $\sqrt{\left\vert T_{2}\right\vert }$ & $\sqrt{%
\left\vert T_{2}\right\vert }$ \\ \hline
$l\neq 0$, $\mathcal{L}_{2}l>0$ &  & finite \\ \hline
$l\neq 0$, $\mathcal{L}_{2}l<0$ &  & $\frac{1}{T_{2}^{2}}$ \\ \hline
$l\neq 0$, $\mathcal{L}_{2}=0$ &  & $\frac{1}{\left\vert T_{2}\right\vert }$
\\ \hline
\end{tabular}

Table 3. Behavior of $\frac{\omega _{2}}{\omega _{1}}$ for the cases when a
photon is emitted near the horizon and is absorbed near the singularity.

Thus we see that there is no qualitative difference between cases with $%
p_{1}=0$ and $p_{1}\neq 0$, if $l=0$. If $p_{1}>0$, the case $l\neq 0$
cannot be realized for emission near the horizon (corresponding cells in
Table 3 are left empty), as is explained above. Meanwhile, it is quite
possible for $p_{1}=0$.

\section{Comparison of the two main scenarios}

Now, having revealed the main features of scenarios i and ii, it is
instructive to compare them. Let us take, for definiteness, the particular
case for which $\mathcal{L}_{2}\neq 0$ and an observer receives radially
propagating signals, $l=0$. The results of comparison based on previous
sections, are collected in Table 4. Here, as before, by "typical" we mean
that points 1 and 2 lie somewhere on the horizon, with $V_{1}$ and $V_{2}$
having the same order. Scenario ii is called special, if $\frac{V_{2}}{V_{1}}%
\gg 1$.

\begin{tabular}{|l|l|l|l|}
\hline
Scenario & Vicinity of horizon & Intermediate & Singularity \\ \hline
i, ii (typical) & Finite & Finite & Unbounded blueshift \\ \hline
ii (special) & unbounded redshift & unbounded redshift & Unbounded blueshift
\\ \hline
\end{tabular}

Table 4. Properties of $\frac{\omega _{2}}{\omega _{1}}$ for different types
of scenario with $l=0$, $\mathcal{L}_{2}\neq 0$.

Thus we see that there is no qualitative difference between scenarios i and
ii (typical). But there is an essential difference between subcases ii
(typical) and ii (special). The combination $\mathcal{L}_{2}\neq 0$, $l=0$
is chosen because the difference under discussion is now especially
pronounced. In this sense, this case is especially interesting.

\section{Summary and conclusions}

Thus the results for absorption of light near the singularity are
qualitatively different in the pure radial and nonradial cases. And, the
change of frequency during a fall of an observer can unfold in different
ways for typical and special scenarios. We hope that account for nonradial
motion carried out in the present paper will be useful further in
investigation of the view of an infalling observer. This should include not
only the description of signals from a remote world outside the horizon but
also view of a close vicinity of an observer himself.

A separate question is the properties of light absorbed near the event \
horizon in the context of the instability issue. We saw that, according to
Table 1, there are cases when (almost) infinite blueshift occurs. If $p=0$,
this is in agreement with the results of \cite{zero}. If $pq<0$ with an
arbitrary $q<0$, this corresponds to a photon coming to the inner black hole
region from the "mirror" universe that is absent in the realistic collapse.
A similar effect occurs near the white hole horizon \cite{white} when a
particle emerging from the white hole horizon collides with another one
moving towards a black hole one. A question arises, whether this can lead to
some new instabilities for the completely extended space-time? The
corresponding conditions of infinite blueshift near the horizon inside
require special fine-tuning (almost zero momentum and trajectories near the
bifurcation point \cite{inner}), so it seems that this constitutes a zero
measure set and does not change an overall picture radically. In a sense,
the same applies to rotating black holes where the BSW effect needs one of
colliding particles to be fine-tuned \cite{ban}. For white holes, this is
not necessary but their instability is already known (see, e.g. \cite{fn}
and literature therein). However, this issue remains highly nontrivial and
requires further investigations in connection with similar effect near inner
horizons.

\section{Acknowledgement}

I thank Alexey Toporensky for helpful discussions. I am also grateful to the
anonymous referee for constructive criticism and useful suggestions. This
work is performed according to the Russian Government Program of Competitive
Growth of Kazan Federal University.

\end{document}